\documentclass{PoS}
\usepackage{verbatim,amsmath,amssymb}
\newcommand{\ba}{\begin{eqnarray}}  
\newcommand{\ea}{\end{eqnarray}}  
\newcommand{\be}{\begin{equation}}  
\newcommand{\ee}{\end{equation}}

\newcommand{\aspi}[1]{\left ( \frac{\alpha_s(#1)}{4\pi} \right ) }

\def\lqcd{\Lambda_{\rm QCD}}

\newcommand{\nn}{\nonumber}

\newcommand{\plus}{\ensuremath{\! + \!}}
\newcommand{\minus}{\ensuremath{\! - \!}}

\newcommand{\lrb}[1]{\left ( #1 \right )}

\newcommand{\lrs}[1]{\left [ #1 \right ]}

\def\lqcd{\Lambda_{\rm QCD}}
\newcommand{\MSb}{\overline{\rm MS}}
\newcommand{\as}{\alpha_s(R)}

\title{The R-evolution of QCD matrix elements}

\ShortTitle{The R-evolution of QCD matrix elements}

\author{Andr\'e H. Hoang, \\ Max-Planck-Institut für Physik (Werner-Heisenberg-Institut), Föhringer Ring 6,\\
80805 München, Germany
E-mail: \email{ahoang@mppmu.mpg.de}}

\author{Ambar Jain\\Center for Theoretical Physics, Massachusetts Institute of
  Technology, Cambridge, MA 02139, USA\\ E-mail:\email{ambarj@mit.edu}}

\author{\speaker{Ignazio Scimemi}
\\
Departamento de F\`\i sica Te\`orica II, Universidad Complutense de Madrid, 28040 Madrid, Spain\\
        E-mail: \email{ignazios@fis.ucm.es}}

\author{Iain W.~Stewart\\
Center for Theoretical Physics, Massachusetts Institute of
  Technology, Cambridge, MA 02139, USA\\
  E-mail:\email{iains@mit.edu}}

\abstract{Perturbation series in QCD are generally asymptotic and suffer from
  so-called infrared renormalon ambiguities. In the context of the standard operator
  product expansion in $\overline{\rm MS}$ these ambiguities are compensated by
  matrix elements of higher dimension operators, but the procedure can be
  difficult to control due to large numerical cancellations.  Explicit
  subtractions for matrix elements and coefficients, depending on a subtraction scale $R$, can
  avoid this problem. The appropriate choice for $R$ in the Wilson coefficients can 
  widely vary for different processes. In this talk we discuss renormalization
  group evolution with the scale $R$, and show that it sums large logarithms in
  the difference of processes with widely different $R$'s. We
  also show that the solution of the $R$-evolution equations can be used to
  recover the all order asymptotic form of the singularities in the Borel
  transform of the perturbative series. For the normalization of these
  singularities we obtain a quickly converging sum rule that only needs the
  known perturbative coefficients as an input. This sum rule can be used as a
  novel test for renormalon ambiguities without replying on the large-$\beta_0$
  approximation.  \\
  
  Preprint numbers:  MIT CTP-4040, MPP-2009-59}

\FullConference{International Workshop on Effective Field Theories: from the Pion to the Upsilon \\
                 1-6 February 2009\\
                 Valencia, Spain}

\begin{document}


\section{The $R$-Evolution}

A well known issue in QCD is that perturbative series in $\alpha_s$ do not
converge, being asymptotic. In perturbative calculations, this feature is
caused by contributions from small loop momenta that are factorially enhanced
due to running coupling effects. Thus an important task in high order QCD
perturbation theory is to establish to which order the perturbative series can
be considered ``meaningful'', or whether it is possible to define subtraction
schemes that improve the convergence properties.  A common instrument to
quantify the high order behavior of perturbative series is the Borel formalism:
poles in the Borel transform of the perturbative series in the strong coupling
$\alpha_s$ correspond to ambiguities that scale like $\Lambda_{\rm QCD}$ to some
integer power $p$. Predictions for physical quantities in
perturbative QCD are rendered ambiguity-free in the framework of the operator
product expansion (OPE)~\cite{Mueller:1984vh}, sometimes formulated within an 
effective theory (EFT) context. Here renormalon ambiguities of Wilson coefficients 
are compensated by matrix elements of higher dimensional operators~\cite{Luke:1994xd}.  
The most common method to determine poles and the corresponding residues in 
the Borel transform is based on the computation of massless quark bubble graphs 
inserted in gluon lines and the ``naive non-abelianization approximation''~\cite{Beneke:1998ui},
also called the ``large-$n_f$'' approximation, which seems to work reasonably
well in many examples.

In an OPE, short and long wavelength contributions are separated into Wilson
coefficients that are computed perturbatively, and matrix elements of operators
that require non-perturbative methods. The most common scheme used in the
literature for this separation of contributions is the $\overline{\rm
  MS}$-scheme. While the $\overline{\rm MS}$ Wilson coefficients are free from
explicit IR divergences to all orders in $\alpha_s$, they may still be sensitive to IR
physics since they incorporate arbitrarily small loop momenta, leading to
renormalon ambiguities. The resulting numerical cancellations between Wilson
coefficients and higher dimensional matrix elements are sometimes difficult to
control reliably for predictions that require small theoretical uncertainties.
This problem can be avoided if schemes for Wilson coefficients and matrix elements
with explicit infrared subtractions are employed.  In general such subtractions
introduce an additional momentum (cutoff) scale, which we denote as $R$. The
choice for the scale $R$ depends on the typical distance scale relevant for the
quantity described in the OPE expansion. It can therefore differ widely for
different applications.

For heavy quark masses such $R$-dependent subtractions schemes have been
employed for more than a decade to avoid the ${\cal O}(\Lambda_{\rm QCD})$
renormalon ambiguity contained in the pole mass~\cite{Beneke:1998ui} and to
define short-distance mass schemes that can be determined with uncertainties
smaller than $\Lambda_{\rm QCD}$.  Depending on the
process the proper choices for $R$ range from about $1$--$2$~GeV in B-physics or
for the description of top jets in the resonance region~\cite{Fleming:2007qr,Jain:2008gb}, to
$15-20$~GeV for top pair production at threshold~\cite{Hoang:2000yr}, to scales
$\gg m$ in cases there the $\overline{\rm MS}$ mass is the proper mass scheme.
For OPE predictions such subtractions have (apart for the just-mentioned heavy
quark masses) only become fashionable recently, relevant  e.g. for a ambiguity-free
definition of the kinetic energy operator $\lambda_1$ in the context of HQET~\cite{Bigi:1994ga},
or for an ambiguity-free definition of the gap-parameter that governs the first
moment of the soft function describing soft radiation in jet
production~\cite{Hoang:2008fs}.

In this talk we discuss the $R$-evolution of matrix elements in schemes with
$R$-dependent subtractions. Much like $\mu$, changes in $R$ are related to changes in 
renormalization scheme.  Due to a power-like dependence on $R$, the solutions
are substantially different from the common well-known logarithmic
renormalization group running.   We show that the corresponding $R$-RGE sums large
logarithms contained in the difference of two subtractions involving
subtractions scales $R_0$ and $R_1$ with $R_0\ll R_1$~\cite{Hoang:2008yj}. This resolves e.g. the
problem of large logs in the relation of low-scale threshold heavy quark masses with $R\ll
m_q$ and the $\overline{\rm MS}$-mass with $R\approx m_q$. Moreover, taking the
formal limit $R_0\to 0$ in the R-RGE solution, we can recover the all order
asymptotic behavior of the singularity in the Borel transform that is associated
to the subtraction.  For the residue of the singularity we obtain a rapidly
convergent sum rule that can be determined from the series coefficients of the
subtractions without relying on the fermion bubble approximation. This sum rule
represents a new method to detect renormalons, and even works for renormalons
dominated by non-abelian effects which are hard to detect with fermion bubbles and 
naive non-abelianization. In this talk we concentrate mostly on the case $p=1$, e.g.  relevant for defining heavy 
quark masses that are free of the ${\cal O}(\Lambda_{\rm QCD})$ renormalon. But we 
emphasize that our method can be applied in a straightforward way for cases with $p>1$ 
as well. For further details we refer to Refs.~\cite{Hoang:2008yj} and~\cite{inprogress}.

Consider a low energy non-perturbative matrix element $\theta_0$ with mass
dimension $p$, which occurs in an OPE for some physical quantity in the
${\overline {\rm MS}}$ scheme and has a leading renormalon ambiguity of ${\cal O}(
\Lambda_{\rm QCD}^p)$. We can write $\theta_0$ as a sum of a matrix element
$\theta(R)$ that is free of the ${\cal O}(\Lambda_{\rm QCD}^p)$ renormalon and a
perturbation series (the scheme dependent subtraction) $R^p \delta\theta(R)$,
which contains the ${\cal O}(\Lambda_{\rm QCD}^p)$ renormalon and introduces the
subtraction scale $R$. We have
\begin{equation}
\label{poleeqn}
\theta_0 = \theta(R) + R^p \delta\theta(R),
\, \quad 
\delta \theta(R)  =   \sum_{n=1}^\infty   a_{n} \,
\Big(\frac{\alpha_s(R)}{4\pi}\Big)^n
\,,
\end{equation}
where the $a_n$ are constants.
Since $\theta_0$ is $R$-independent it is straightforward to determine the
$R$-evolution equation for the $\theta(R)$, and corresponding $R$-anomalous
dimension $\gamma[\alpha_s(R)]$,
\begin{equation}\label{R-RGE}
R\frac{d}{dR}\theta(R) =  - R^p \gamma[\alpha_s(R)]\, ,
\, \quad \gamma [\alpha_s(R)] = \sum_{n=0}^{\infty} \gamma_n\, \left (
  \frac{\as}{4\pi} \right )^n
\,.
\end{equation}
Here the coefficients are $\gamma_{n-1} = p\, a_n - 2 \sum_{k=1}^{n-1} k\, a_k\, \beta_{n-k-1}$. 
Since $R d/dR \Lambda_{\rm QCD}^p =0$, the series for $\gamma[\alpha_s(R)]$ is
free of the ${\cal O}(\Lambda_{\rm QCD}^p)$ renormalon. We can extend the
situation to the more general case where the matrix element is renormalization
scale dependent, $\theta_0 = \theta_0(\mu)$. In this case the subtraction is
also generally renormalization scale dependent, leading to a matrix element
$\theta(\mu,R)$ in an $R$-subtraction scheme. Here $\theta(\mu,R)$ satisfies
both a renormalization group equation in $\mu$, and also an $R$-evolution
equation for $\mu=R$ that has as a form identical to eqn.(\ref{R-RGE}). We note
that in the $\mu$-dependent situation, it can be quite subtle to construct a
subtraction that leads to consistent evolution equations in $\mu$ and $R$.
Examples where this has been achieved are the top quark jet mass that is relevant
for describing the single top invariant mass distribution in top
production~\cite{Jain:2008gb}, and the gap parameter in the soft function
governing dijet production in $e^+e^-$ annihilation~\cite{Hoang:2008fs}.


\section{General Solution of the R-RGE}

In order to solve eqn.~(\ref{R-RGE}) we need an all order solution of the
$\alpha_s$-RGE.  Consider the $\alpha_s$-RGE,
$
 R \frac{d \as}{dR} = -\alpha_s^2(R)/(2 \pi)\: \sum_{n=0}^{\infty}\beta_n \lrb{ \as/4\pi}^n \, ,
$
where $\beta_n$ are $\beta$-function coefficients in an arbitrary scheme for the strong coupling.
A solution of this equation is given by
\begin{eqnarray}\label{alphasRGE}
 \ln\frac{R_1}{R_0} =\!\! \int_{\alpha_0}^{\alpha_1}\!\! \frac{d\alpha_R}{\beta[\alpha_R]} 
  =\! \int_{t_1}^{t_0}\!\!\! dt\: \hat b(t) = G(t_0) - G(t_1) \,,
\end{eqnarray}
where $\alpha_i\equiv \alpha_s(R_i)$, $\alpha_R\equiv \alpha_s(R)$, $t_i \equiv
-2\pi/(\beta_0\alpha_i)$, and $t\equiv -2\pi/(\beta_0\alpha_R)$.  Here
\begin{equation}
\hat b(t) = 1\plus \frac{\hat b_1}{t} \plus \frac{\hat b_2}{t^2} \plus \frac{\hat b_3}{t^3} \plus  \ldots \, , ~~~~~~~~
G(t) = t \plus \hat b_1 \ln(-t) \minus \frac{\hat b_2}{t} \minus \frac{\hat b_3}{2t^2} \minus \ldots \, ,
\end{equation}
where for the first few orders $\hat b_1 = \beta_1/(2\beta_0^2), ~ \hat b_2 =
(\beta_1^2 \minus \beta_0\beta_2)/(4\beta_0^4)$, and $\hat b_3 = (\beta_1^3
\minus 2\beta_0\beta_1\beta_2 \plus \beta_0^2\beta_3)/(8\beta_0^6)$.  For later
convenience we also define
$G_2(t) = G(t) - t - \hat b_1 \ln(-t) \, .$
From eqn.~(\ref{alphasRGE}) one immediately notes that, $R\, \exp{G(t)} = R_0\,
\exp{G(t_0)} \equiv \lqcd$.  This equality demands $R e^{G(t)}$ to be a constant
of mass dimension one, which we have defined to be $\lqcd$ 
and one can easily check that this definition corresponds to the familiar
definition of $\Lambda^{(k)}_{\rm QCD}$ at ${\rm N}^k{\rm LL}$ order. The
solution above is valid for an arbitrary mass-independent scheme for the $\beta$-function, although
usually one uses the $\MSb$ scheme where the $\beta_i$ are known to
four-loops~\cite{vanRitbergen:1997va}. Here we present the solution for the
$R$-RGE of eqn.~(\ref{R-RGE}) for $p=1$. For arbitrary $p$, see
\cite{inprogress}.
Integrating equation~(\ref{R-RGE}) gives,
\begin{eqnarray}\label{sol1}
\theta(R_1) - \theta(R_0) &=& - \int_{R_0}^{R_1} dR \,  \gamma[\as]  
 = \int_{t_0}^{t_1} dt \, \hat b(t) \lrs{\lqcd \, e^{-G(t)}}  \gamma[t]  \, ,
\end{eqnarray}
where we have used the solution of the $\alpha_s$-RGE and defined $\gamma[t]
\equiv \gamma[\as]$ using the substitution $t = -2\pi/(\beta_0\as)$.  In order
to write the solution in a closed form we define
\begin{eqnarray}\label{Sjp}
&& \hat b(t)\,
e^{- G_2(t) } \,  \gamma[t]  \equiv \sum_{j=0}^\infty  S_j  
(-t)^{-j-1}    \, , \text{ where}\; \;  
 S_0  =  \frac{\gamma_0}{2\beta_0} \,, 
  \quad
 S_1  =  \frac{\gamma_1}{(2\beta_0)^2} - (\hat b_1\plus \hat b_2)\frac{\gamma_0}{2\beta_0}\,,
 \nn  \\
 &&\ \  S_2  = \frac{\gamma_2}{(2\beta_0)^3} - (\hat b_1\plus \hat b_2) \frac{ \gamma_1}{(2\beta_0)^2}
    +\big[ (1\plus \hat b_1)\hat b_2 +(\hat b_2^2 \plus \hat b_3)/2 \big]
  \frac{ \gamma_0}{(2\beta_0)} , \ldots \,.
\end{eqnarray}
Now substituting eqn.~(\ref{Sjp}) into eqn.~(\ref{sol1}) and using the definition
of the incomplete gamma function with the standard convention of integration
above the cut, $\int_{t}^{\infty} dt' \, (-t')^{c-1} e^{-t'} = - e^{-i\pi c} \,
\Gamma(c,t)$, we get the N$^k$LL solution that is given by
 \begin{eqnarray}\label{solNkLL}
 [\theta(R_1) - \theta(R_0)]^{{\rm N}^k{\rm LL}} &=& (\lqcd^{(k)}) \sum_{j=0}^{k} S_j  (-1)^j
  e^{i \pi \hat b_1} 
   \lrs{\Gamma(-j- \hat b_1,  t_1) - \Gamma(-j-\hat b_1,  t_0)} \, .
 \end{eqnarray}
Here $\lqcd^{(k)}$ is the N$^k$LL solution of: $ R\, \exp{G(t)} \equiv \lqcd.  $
The integrand in eqn.~(\ref{sol1}) has an essential singularity and a branch
point at $t=0$ due to infinitely many negative powers of $t$ and a branch cut on
the positive real axis due to the $(-t)^{-\hat b_1}$. For $R_0, R_1 > \lqcd$ we have
$t_0, t_1 < 0$, so the range of integration is away from the branch cut and the
integral is well defined. To write a closed form expression we have split the
integral into a difference of two incomplete gamma functions, and with the
$e^{i\pi\hat b_1}$ factor this solution is real.  
 Note that only the difference of the gamma functions appearing in
 eq.~(\ref{solNkLL}) is independent of the convention for treating the cut.

To see what kind of
perturbative terms the solution in eq.~(\ref{solNkLL}) contains, we take the $k=0$ case (LL) and  recall the
asymptotic expansion for $t\to -\infty$,
 $\Gamma[c,t] \stackrel{\rm asym}{=} 
  e^{-t}\, t^{c-1} \sum_{n=0}^\infty 
  \frac{\Gamma(1\minus c\plus n)}{(-t)^n\, \Gamma(1\minus c)}
$  \,.
For $c=0$ this yields $ \Lambda_{\rm QCD}^{(0)} \Gamma[0,t] \stackrel{\rm
  asym}{=} -2R \sum_{n=0}^\infty 2^n\, n!\,
\Big(\beta_0\alpha_s(R)/(4\pi)\Big)^{n+1}$, where we used the LL relation
$\Lambda^{(0)}_{\rm QCD}e^{-t}=R$.  This is a divergent series, but for the LL
solution of the RGE
\begin{eqnarray}\label{expnmm}
[  \theta(R_1)-\theta(R_0) ]^{\rm LL}
  &=& \frac{-\gamma_0 R_1 }{2\beta_0} \sum_{n=0}^\infty 
   \Big[\frac{\beta_0\alpha_1}{2\pi}\Big]^{n+1}  n!\, \bigg( 1 \!-\! \frac{R_0}{R_1} 
   \sum_{k=0}^n \frac{1}{k!} \ln^k\!\frac{R_1}{R_0}\bigg) 
  \nn \\
  & =& -\frac{\gamma_0 R_0 }{2\beta_0}  \sum_{n=0}^\infty 
   \Big[\frac{\beta_0\alpha_1}{2\pi}\Big]^{n+1}  
   \sum_{k=n+1}^\infty \frac{n!}{k!} \ln^k\!\frac{R_1}{R_0} \,,
\end{eqnarray}
which is convergent since $\beta_0\alpha_s(R_1)\ln(R_1/R_0)/(2\pi) <1$.
Eqn.~(\ref{expnmm}) displays the problem of large logs in fixed order
perturbation theory for $R_1\gg R_0$.  The RGE in $R$ encodes IR physics from
the large order behavior of perturbation theory.  It causes a rearrangement of
the IR fluctuations included in $\theta(R)$ in going from $R_0$ to $R_1$ without
reintroducing a renormalon.

\section{The R-RGE and the Leading Renormalon Ambiguity}

\label{sec:amb}

In this section we demonstrate a novel application of the $R$-RGE by deriving the leading renormalon ambiguity of $\theta_0$ of eqn.~(\ref{poleeqn}) in the Borel plane and providing a sum rule to perturbatively determine the normalization of the renormalon ambiguity. A sum rule of this kind for $u=1/2$ ambiguity of the pole mass was first provided in our earlier work \cite{Hoang:2008yj}. 

We note\footnote{Eqn.~(\ref{alphasRGE}) has the solution that as $t$ goes to
  infinity, $R \to 0$ with a phase $e^{\pm i \pi \hat b_1}$. That is to say $R$
  cannot approach $0$ from the positive real axis. $R$ must go to $0$ either from
  above the positive real axis or below it. This is so because there is a cut on the
  positive real axis for $0 < R < \lqcd$. This cut corresponds to the branch cut
  in the $t$-plane for the integrand of the equation~(\ref{sol1}). Nevertheless,
  $t$ as a function of $R$ is single-valued at $R=0$, corresponding to
  $\alpha_s(R=0)=0$.}  that in the complex $R$-plane away from the positive real
axis, $\lim_{R \to 0}\alpha_s(R) = 0$. Therefore $\delta \theta(R)$ of
eqn.~(\ref{poleeqn}) vanishes at $R=0$ and we get $
\lim_{R \to 0}\theta(R) = \theta_0 \, .  $ 
It is important to note that though $\alpha_s(R=0)$ is single-valued, but the
process of taking the limit $R \to 0$ is not unambiguous due to the branch cut
on $0<R<\lqcd$.  It is the limiting procedure that reintroduces the ambiguity.
Thus, taking limit $R_0 \to 0$ in eqn.~(\ref{poleeqn}) we get
 \begin{eqnarray}\label{resummed-pole-eqn}
  \theta(R_1) - \theta_0     
 & =& - \lqcd \sum_{j=0}^{\infty} S_j \,   \int_{ t_1}^\infty \! dt  \frac{ e^{-t}}{(-t)^{j+\hat b_1+1}} \, . 
 \end{eqnarray}
 The LHS has a renormalon ambiguity in $\theta_0$, which on the RHS is in the
 integral, the integrand has a branch cut on the positive real axis in the
 complex $t$-plane, and the integration must be performed by either going above
 the positive real axis or below it. Using the asymptotic expansion for this
 integral, which is the same as that for $\Gamma(c,t)$, 
 and  expanding 
$ e^{ G_2(t)} = \sum_{\ell =0}^{\infty} \frac{g_\ell}{(-t)^\ell}  \, , $
we get a power series expansion in $1/(-t) = \beta_0 \alpha_s(R)/(2\pi)$,
\begin{eqnarray}
 \theta(R) - \theta_0 &=&  - R \ \sum_{j,n,\ell=0}^{\infty} S_j g_\ell  \frac{\Gamma(1+j+n+ \hat b_1 )}{\Gamma(1+j+\hat b_1)(-t)^{n+j+\ell+1}}  \, .\nn
\end{eqnarray}
On taking the Borel transform\footnote{Taking the Borel transform amounts to making the replacement $1/t^{n+1} \to 2(2u)^n/\Gamma(n+1)$. In our convention, which is also the most widely used, $u$ is Borel conjugate of $\beta_0\alpha_s/(4\pi)$.}, after some algebra, we get
\begin{eqnarray} \label{Bpp}
  B(u) =  2 R \bigg[ \sum_{\ell=0}^\infty g_\ell\,
 Q_\ell(u)  
   -  P_{1/2} \sum_{\ell=0}^\infty g_\ell\,\frac{ \Gamma(1\plus  \hat
 b_1 \minus \ell) }{(1\minus 2u)^{1+  \hat b_1-\ell}} \bigg], 
\end{eqnarray}
where
\begin{eqnarray}
\label{sumrule}
P_{1/2}  &=& \lim_{k \to \infty} P_{1/2}^{\rm N^kLL} = \lim_{k \to \infty} \sum_{j=0}^k  
\frac{ S_j }{\Gamma(1\plus \hat b_1\plus j)} 
\end{eqnarray}
and $Q(u)$ is a function convergent in a finite neighborhood of $u=1/2$.
The difference, $\theta(R)-\theta_0$, can be obtained by the inverse Borel
transform, $
\theta(R) -\theta_0 = \int_0^\infty \!\! du \ B(u)\ e^{-u \
  4\pi/(\beta_0\alpha_s(R))} \, .  $ 
This integral is ambiguous because eqn.~(\ref{Bpp}) has a branch cut for
$u>1/2$. Thus we have obtained, in eqn.~(\ref{Bpp}), the structure of the
leading renormalon ambiguity in $\theta_0$, represented by the second term on
the RHS, along with its normalization $P_{1/2}$.  Here $P_{1/2}$ is an
absolutely convergent infinite series~\cite{Hoang:2008yj} constructed from the
coefficients $S_j$ that appear in the solution of the R-RGE (eqn.~(\ref{Sjp})).
The $S_j$ are in turn determined by the $R$-anomalous dimension coefficients
$\gamma_n$ (eqn.~(\ref{R-RGE})).  Therefore $P^{{\rm N}^k{\rm LL}}_{1/2}$ is
perturbatively determinable in terms of the original coefficients $a_n$ of
$\delta \theta(R)$.

For any given $\theta(R)$ the process described above can be repeated with a
rescaling $R \to \lambda R$. This does not change the normalization of the
renormalon ambiguity, since it amounts to only changing the scale $R$, or
equivalently changing the scheme. In this new $\lambda$-dependent scheme,
we have $\lambda$-dependent $S_j$, where the first few are given by
\begin{eqnarray}
S_0 (\lambda) &=&   \lambda S_0\,, 
  \quad
 S_1  (\lambda)= \lambda S_1 -\frac{a_1 \lambda  \log (\lambda )}{2\beta_0}\ ,
\nn \\
 S_2  (\lambda)&=&\lambda S_2+\frac{\lambda  \log (\lambda ) (-a_2+a_1 (\hat{b}_2+2) \beta_0+a_1 \beta_0
   \log (\lambda ))}{2 \beta_0^2} \, .
\label{eq:sjlambda}
\end{eqnarray}
As a result $P_{1/2}^{\rm N^kLL}$ depends on $\lambda$, but $P_{1/2}$ will be
$\lambda$-independent. This gives a way to assign errors and test for
convergence of the partial sum $P_{1/2}^{\rm N^kLL}$. Distinguishing $P_{1/2}\ne
0$ and $P_{1/2}=0$ provides a test for the presence of a renormalon ambiguity.
We propose to keep $0.5 < \lambda < 2$, so that the perturbative expansion is
well behaved.



\section{Applications of the Renormalon Sum Rule}\label{sec:app}

\begin{figure}[t]
\begin{center} \includegraphics[height= \columnwidth,angle=-90]{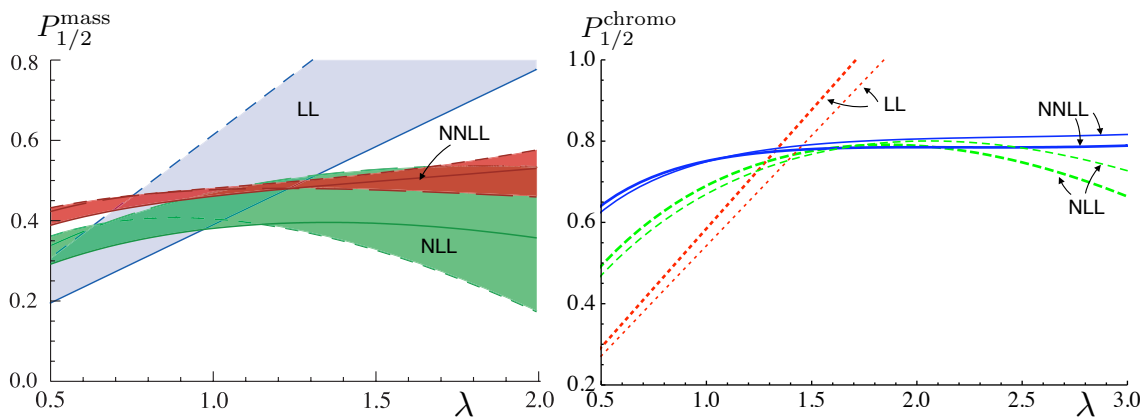} \end{center}
\vskip-0.6cm
\caption{The $P_{1/2}$ sum rule for  the heavy quark pole mass (left) and  the chromomagnetic operator Wilson coefficient (right).}
\label{fig:pole-mass-sum-rule}
\end{figure}

As an application of the sum rule, we calculate the residue of the renormalon
ambiguity of the heavy quark pole mass. Using eqn.~(\ref{sumrule}) along with
$S_j$'s from eqn.~(\ref{eq:sjlambda}), we obtain $P_{1/2}$ as a function of
scaling $\lambda$ in MSR \cite{Hoang:2008yj,Hoang:2008xm,inprogress}, static
\cite{Hoang:2008yj,inprogress} and PS \cite{Beneke:1998ui} mass schemes up to
NNLL order. This is shown in Figure~\ref{fig:pole-mass-sum-rule}, where the band
that envelops $P_{1/2}$ curves for these schemes (with $n_f = 5$ light flavors) is
plotted at LL, NLL and NNLL order. Since the pole mass renormalon ambiguity
is scheme independent, we expect the sum rule to converge to a scheme
independent number. This is clearly seen from the figure, since the band at the
NNLL order is significantly narrower with weak $\lambda$-dependence.  From the
narrow red (NNLL) band we estimate that $P_{1/2}^{mass} = 0.47 \pm 0.10$, which is the
normalization of the order $\lqcd$ renormalon ambiguity in the heavy quark pole
mass. For comparison, the bubble chain approximation gives $P_{1/2}^{mass} = 0.80$,
overestimating by a factor of 2. Our values are consistent with the determination of the normalization of the renormalon ambiguity in Ref.~\cite{Pineda:2001zq}.

The Wilson coefficient $C_{\rm cm}(m_Q,\mu)$ of the chromomagnetic HQET operator
$\bar h_v \sigma\cdot G h_v$, is known to have a ${\cal O}(\lqcd/m_Q)$
renormalon.  We will test for this renormalon using our renormalon sum rule and
determine its size.  $C_{\rm cm}$ is known to three loops~\cite{Grozins} and at
$\mu = m_Q$, it is given by, $C_{\rm cm}(m_Q, m_Q) -1 = \sum_{i=1}^\infty a_i
\aspi{m_Q}^i$.
We are looking for $n!(2\beta_0)^n$ growth ($p=1$) in the coefficients of
$\as^n(m_Q)$. To test this, we take the known $\overline {\rm MS}$ scheme
results $ a_1 = 8.66703$, $a_2 = (350.347 - 30.6037 n_f)$, $a_3 = (21985.2 -
3470.72 n_f + 101.8 n_f^2)$~\cite{Grozins} and substitute in
eqn.~(\ref{eq:sjlambda}), which we then use in eqn.~(\ref{sumrule}) to get the
$P_{1/2}$ sum rule as a function of $\lambda$.  This sum rule is plotted in
Figure~\ref{fig:pole-mass-sum-rule} for $n_f = 3$ (thin lines) and $n_f=4$
(thick lines) light flavors at the LL(dotted), NLL(dashed) and NNLL(solid)
orders; convergence is quite evident and from the $\lambda$ dependence we estimate
$P_{1/2}^{chromo} = 0.72 \pm 0.09$ ($n_f = 3$), and $P_{1/2}^{chromo} = 0.71 \pm 0.07$ ($n_f = 4$).
The sum rule yields a clear renormalon.
%
For further details and discussions
regarding concepts presented in this proceeding, including some direct
applications of the $R$-RGE, see~\cite{Hoang:2008yj,inprogress}.

{\bf Acknowledgments:} This work was supported by the EU network contract MRTN-CT-2006-035482
(Flavianet), and the Office of Nuclear Physics of the U.S.\ Department of
Energy, contract DE-FG02-94ER40818.

%


\end{document}